\documentclass[journal,transmag]{IEEEtran}
\ifCLASSINFOpdf
\else
\fi

\usepackage{graphicx}
\usepackage{float}
\usepackage{amsmath}
\usepackage{amssymb}
\usepackage{graphicx}
\usepackage{hyperref}
\usepackage{color}
\usepackage{booktabs}
\usepackage{subfigure}
\usepackage{algorithm}
\usepackage{algorithmic}

\begin{document}
%

\title{SFCaaS: Service Function Chains as a Service in NFV Environments}


\author{\IEEEauthorblockN{Tarik Moufakir\IEEEauthorrefmark{1},
Mohamed~Faten~Zhani\IEEEauthorrefmark{1},
Abdelouahed~Gherbi\IEEEauthorrefmark{1}, 
Moayad Aloqaily\IEEEauthorrefmark{2}, and
Nadir Ghrada\IEEEauthorrefmark{1}}
\IEEEauthorblockA{\IEEEauthorrefmark{1}Ecole de Technologie Sup\'erieure (\'ETS),
Montreal, Quebec, Canada}
\IEEEauthorblockA{\IEEEauthorrefmark{2}Machine Learning Department, Mohamed Bin Zayed University of Artificial Intelligence (MBZUAI), UAE
}
\thanks{
Corresponding author:M. F. Zhani (email: mfzhani@etsmtl.ca}
}

%



\maketitle

\begin{abstract}
\textit{Abstact---}With the emergence of network softwarization trend, traditional networking services offered by Internet providers are expected to evolve by fully leveraging new recent technologies like network function virtualization and software defined networking. In this paper, we investigate offering Service Function Chains as a Service (SFCaaS) in NFV Environments. We first describe the potential business model to offer such a service. We  then conduct  a  detailed  study  of  the  costs  of  virtual  machine instances offered by Amazon EC2 with respect to the location, instance  size,  and  performance in order to guide service chain provisioning and resource  allocation.  Afterwards, we address the resource allocation problem for service chain functions from the SFC provider's perspective while leveraging the performed cost study. We hence formulate the problem as an Integer Linear Program (ILP) aiming at reducing the SFC provider’s operational costs of virtual machine instances and links as well as the synchronization costs among the instances. We  also  propose  a new  heuristic  algorithm  to  solve  the  mapping  problem with  the  same  aforementioned  goals  taking  into  account  the  conducted study of the costs of Amazon EC2 instances. We show through extensive simulations that the proposed heuristic significantly reduce operational costs compared to a Baseline algorithm inspired by the existing literature.
\end{abstract}

\begin{IEEEkeywords}
SFCaaS, Resource Allocation, Network Softwarization, Service Function Chaining, Network Function Virtualization, FlexNGIA.
\end{IEEEkeywords}



%
\IEEEpeerreviewmaketitle

\section{Introduction}
%
%
%
%
\IEEEPARstart{P}{ropelled} by recent advances in Virtualization technologies,  Network Function Virtualization (NFV), Software Defined Networking (SDN) and data plane programmability,  network softwarization has become a new trend aiming at separting software from the networking hardware. This trend is revolutionizing the way networks are designed and managed and opening the door for a new IT open ecosystem with new plartforms, software and network applications \cite{FlexNGIA2019,clemm2020network,Varghese2019}.
%
As a matter of fact, network softwarization benefits are numerous including better agility and flexibility, easy automation and faster time-to-Market services while ensuring their scalability and significantly reducing capital and operational expenditures.

\begin{figure}
	\begin{center} 
	\includegraphics[scale=.30]{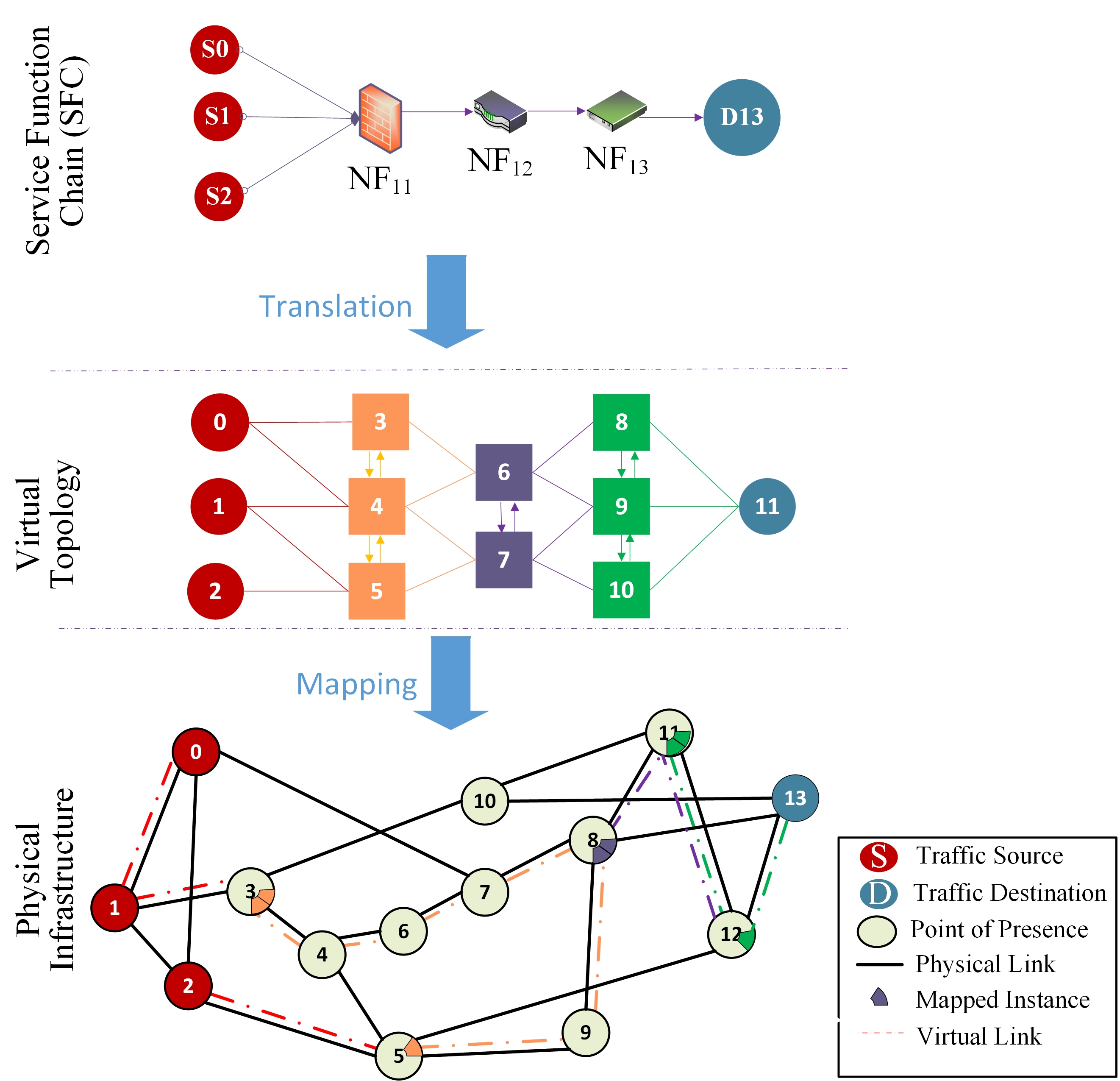}
	\end{center}
	\caption{SFCaaS - SFC Translation and Mapping}
	\label{Fig:SFCTranslationMapping}
\end{figure}

Network Softwarization allows network/cloud providers (called hereafter SFC provider) to roll out a new service offering, namely “Service Function Chains” as a Service (SFCaaS) where a whole Service Function Chain (SFC) could be offered as a service to a third party (called hereafter service provider). An SFC is an ordered set of Virtual Network Functions (VNF) that are crossed by packets arriving from the chain's sources and flowing towards the chain's destinations~(Fig.~\ref{Fig:SFCTranslationMapping}). VNFs~are network functions (e.g., firewall, IDS or other functions) implemented in virtual machines or containers running on commodity servers or implemented in~dedicated hardware like Field-Programmable Gate Arrays (FPGAs).

In this context, the SFC provider would face a major challenge as to how to allocate the resources to the requested SFC in the physical infrastructure. Although this problem has been recently extensively addressed \cite{racheg2017profit,Tashtarian2019,amokrane2013greenhead} in the literature, we revisit this problem in this work with the following novel contributions:
\begin{enumerate}
\item We consider two phases to provision an SFC (Fig~\ref{Fig:SFCTranslationMapping}). The first one is the translation phase which aims at identifying the optimal number of  instances (i.e.,~virtual machines/containers) that are required for each VNF in order to cater to the demand. The second phase is the mapping phase aiming at deciding where to place the instances taking into account the deployment costs, which signifantly varies from one location to another. To our knowledge no prior work considered the translation phase.
\item We conduct a detailed study of the costs of Virtual Machines (VMs) (i.e.,~instances) offered by Amazon EC2 \cite{AmazonEC2Instances2021} with respect to the location, instance size, and performance. This cost study extends the one carried out by the authors in \cite{Ghrada-PVESDN2018} and shed the light on interesting observations about the costs of instances in one of the major cloud providers in the market. This study is used later on in this paper to guide the resource allocation for the service function chains. 
%
%
\item We formulate the mapping phase problem as an Integer Linear Program (ILP) aiming at reducing the SFC provider's operational costs of the VM instances and links as well as the synchronization costs among the same-type instances. It is also worth noting that Synchronization costs were not considered in existing literature. 
\item We also propose two heuristic algorithms to solve the mapping problem with the same aforementioned goals taking into account the conducted study of the costs of Amazon EC2 intances. The first one is a basic and intuitive mapping algorithm (called Baseline) and the second is a more sophisticated algorithm called SFC compoSition-based ProvisIoNing (SPIN) algorithm. We show through extensive simulations that SPIN significantly reduces the operational costs compared to the Baseline. 
\end{enumerate}
The remaining of this manuscript is organized as follows. 
Section~\ref{Section:BusinessModel} describes the SFCaaS service model, highlighting the business model and the potentional involved stakeholders as well as the technical challenges related to the deployment of such a service.
Section~\ref{sec:RelatedWork} overviews the related literature on the service function chaining highlighting the novelty of this work.
Section~\ref{Section:StudyAWS} presents the detailed study of the costs of Amazon EC2 instances with respect to several parameters and presents its main outcomes.
Section~\ref{Section:ProblemFormulation} presents the ILP formulation of the mapping problem.
Furthermore, Section~\ref{Section:ProposedSolutions} introduces the two proposed greedy algorithms to address the mapping problem.
The experimental results are provided in Section~\ref{Section:PerformanceEvaluation}.
Finally, conclusions and key research directions are described in Section~\ref{Section:Conclusion}.


\section{SFCaaS - Business Model, Benefits and Challenges}\label{Section:BusinessModel}


In this Section, we propose the following business model adapted to SFCaaS services. This model identifies the stakeholders involved in an environment where SFCs are provisioned and offered as a service. We~mainly identify three stakeholders defined as follows:
\begin{itemize}
    \item SFC Provider: it~is~a company that offers “Service  Function Chains” as a Service (SFCaaS). It owns and manages a physical infrastructure and is in charge of deploying platforms and software required to run network functions needed for the chains and also of provisioning and managing the requested SFCs. It should hence ensure the SFC translation phase to generate a virtual network (Fig.~\ref{Fig:SFCTranslationMapping}), composed of virtual machines and links, which should be later on mapped onto the infrastructure.
    The virtual network is obtained by identifying the number of instances needed to implement each VNF and the virtual links used to connect them (Fig.~\ref{Fig:SFCTranslationMapping}). Note that  virtual links should be also created to connect the instances implementing the same VNF in order to ensure synchronization among them. Indeed, synchronization is needed to guarantee the normal operation of some network functions (e.g.,~IDS)   \cite{AlomariCNSM2020}.
    \item Service Provider: it could be a company or Institution that has users spread around the world. A service provider needs to define the SFC needed to run its service, its composition, performance requirements, and identifies the chain sources/destinations. The composition of the SFC refers to the type of each of the network functions (NFs) making up the chain. The performance requirements could be in terms of end-to-end delay, packet loss, traffic demand, resource (cpu, memory and disk) and other parameters. Of course, the service provider relies on the SFC provider to provision the SFC and allocate the needed resources.
    \item User: users are customers of the service provider and are located at the sources or destinations of the service function chain. The traffic coming from users  will be steered across the SFC provisioned by the service provider.
\end{itemize}

\begin{figure}
	\begin{center} 
	\includegraphics[scale=.65]{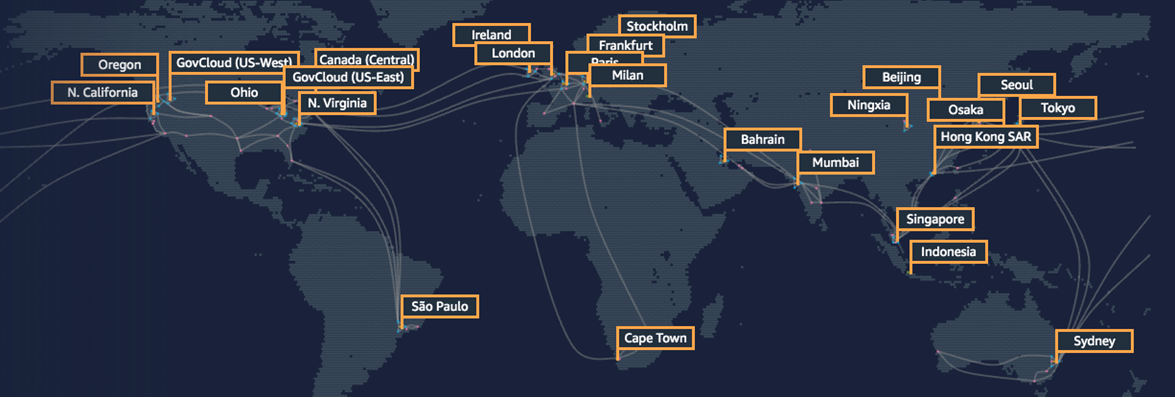}
	\end{center}
	\caption{AWS Global Infrastructure - Source \cite{AWScloud2021}}
	\label{Fig:AWSInfrastucture}
\end{figure}

Potential SFC providers could be major companies offering cloud services like  Google, Amazon EC2, and Microsoft that have their own network with predictable performance \cite{AWScloud2021,googlecloud2021}. 
For instance, the AWS infrastructure shown in Fig.~\ref{Fig:AWSInfrastucture} is a software defined world-wide global Infrastructure \cite{AWScloud2021} with 25~regions across the world serving 245 Countries and territories  where each region contains one or more data centers with 218 Edge Locations and 12 Regional Edge Caches for a total of 230 ~Points of~Presence. Regions are connected through a private world wide network managed by Amazon AWS, which makes it easy to Amazon to predict and even fully control the network performance. 
As computing resources are available everywhere in this global infrastructure and such companies have the expertise on technologies like cloud computing, virtualization, and~SDN, it is straightforward that they could provide SFCs as a Service and easily provision and allocate the required resources across their global infrastructure. 

$\bullet$ \textbf{SFC Benefits:}:
Similar to traditional cloud computing services, offering SFCs a a service would bring several benefits for service providers like avoiding management hassle with no need to software and hardware maintenance. In addition, costs will be reduced as there is no capital and operational expenditures for service providers. It will be also possible to get low prices thanks to the  economies of scale. Furthermore, SFCaaS would allow better performance thanks to the SFC provider’s expertise and knowledge of the infrastructure (i.e., topologies, characteristics, performance). SFC providers could also offer network functions that are carefully implemented and managed to offer optimal performance.

$\bullet$ \textbf{Technical Challenges:}
From the SFC provider's perspective, the main challenge is to provision SFCs while maximizing profit, minimizing operational costs and satisfying the SFC requirements in terms of end-to-end delay. In this context, we can identify several challenges that can be summarized as follows:
\begin{itemize}
    \item Decide how many instances to use to implement a VNF: Implementing a VNF in a single instance (VM/Container) may not be sufficient for several reasons. Indeed, a single instance may not have enough resources to handle the incoming traffic. Hence, implementing the same VNF in multiple instances allows to overcome the lack of resources by distributing the function over several PoPs. It~also allows to reduce costs and to improve fault tolerance as the failure of an instance would not affect the others. At the opposite side, when a VNF is implemented in  multiple instances, there might be some drawbacks. For instance, some data might need to be synchronized among the different instances in order to ensure a normal operation of the network function (e.g.,~distributed Intrusion Detection Systems \cite{AlomariCNSM2020}). As a result, one need to consider the cost of this  synchronization among the same-type instances. This cost can be expressed in terms of CPU, memory and bandwidth consumption to ensure  synchronization. There are might be also constraints on the delay needed to carry out the synchronization. In~this~context, deciding how many instances are needed to implement one VNF and what are the synchronization costs and constraints are key challenges when provisioning the SFC.\\ 
    \item Decide of the type of VM instance to use to run the VNF: The selection of the VM instance type depends on network function requirements in terms of resources (i.e., vCPU, memory, storage),  processing capacity (packets per second),  and the operational cost of running the instance. Of~course, the decision should take into account the network function properties (the nature of the function itself, the used software and operating system, database and other software) as well as the geographical location of the instance which has an impact on the cost and access  delay.\\
    \item Place and chain the instances: the third challenge is to identify where to place the instances in the physical infrastructure and how to allocate the bandwidth resources to chain them in order to mazimize the SFC provider's profits and minimize its operational costs.
\end{itemize}

In this work, we try to approach the aforementioned challenges and study the related parameters and considerations to address them. In the following, we summarize the existing literature that has attempted to address these challenges.

\section{Related Work} \label{sec:RelatedWork}
In this section, we briefly present recent research work addressing the provisioning problem of service function chains.  In the recent years, a large body of work has studied this problem consisting in~the~placement and chaining of~a~set of~ordered VNFs~\cite{ElasticVNF2015,Racheg2017,carpio2017vnf,Wang2016,AlomariCNSM2020}. 
Several research works have aimed to minimize operational costs~\cite{murukan2016cost,zhang2017proactive,gupta2017colap}, to minimize network utilization~\cite{murukan2016cost,ma2017sdn}, to minimize latency \cite{li2019availability,patel2017mobility,yang2016energy,bhamare2017optimal,el2016energy,krishnaswamy2015latency,kim2016energy} and~to~minimize resource consumption~\cite{tajiki2018joint,xu2018energy,addis2015virtual,yang2016energy,el2016energy,bari2016orchestrating,kim2016energy, pham2017traffic}. The analysis of existing literature on SFC placement and~chaining carried out by~Santos~et~al.~\cite{santos2022service} finds that~the~minimization of~operational costs as~the most widely researched goal by 42\% of~published articles.

For instance, Carpio et al \cite{carpio2017vnf} formulate the placement and chaining of VNFs as a mixed linear program and compare it with a random fit placement algorithm. 
For a better scalability, they devise a genetic algorithm to find a sub-optimal solution. The proposed algorithm  focuses first on finding admissible paths and calculate link costs then allocating within those paths the resources for VNFs. However, the algorithm assumes that the number of VNF instances is known beforehand.


 Bari et al. \cite{Bari2015nf.io} designed a platform to carry out the placement, the chain composition and the monitoring of VNFs. This proposal defines the linux file system used and features that can be exploited to implement some certain tasks; however, the proposed method neither addresses the case where multiple instances are instantiated nor determines the appropriate placement of VNFs in the infrastructure.
Beck and Botero \cite{Beck2015GLOBECOM} looked also at the VNF placement and chaining problem and proposed CoordVNF, a solution that  aims at minimizing the link utilization over the infrastructure. This proposal considered the use of multiple instances for the virtualized deep packet inspection function where the traffic is split into TCP and non-TCP traffic. 

%
Wang et al.~\cite{Wang2016} address the problem of online deployment of scalable multiple VNF instances in order to process the fluctuating traffic rate received at the VNFs in order to minimize the cost of the provisioned resources. The authors proposed two algorithms. One for  provisioning a single service chain and the other for provisioning simultaneously multiple service chains.

Ghaznavi et al.~\cite{ElasticVNF2015} address the VNF Placement problem with the goal of reducing server and bandwidth consumption. They introduce a solution to optimize the placement of VNFs by minimizing installation, transportation, reassignment and migration costs of VNF instances. However, the solution assumes that all instances of the chain are of the same type.

Unlike previous work, in this paper, we consider not only SFC mapping but also the translation phase where the number of instances for each VNF is estimated and considered in order to build a virtual network to be mapped. We also address the SFC mapping taking into account synchronization and deployment costs of the VNF instances. Our work relies on realistic data and study based on the Amazon EC2 instance costs.

In the following, we start first by studying the costs of the offered general-purpose instances as this study will guide the development of the proposed solutions for the virtual machine translation and mapping phases. 
\section{Study of the costs of Amazon EC2 instances}\label{Section:StudyAWS}

\begin{figure}
	\begin{center} 
	\includegraphics[scale=.50]{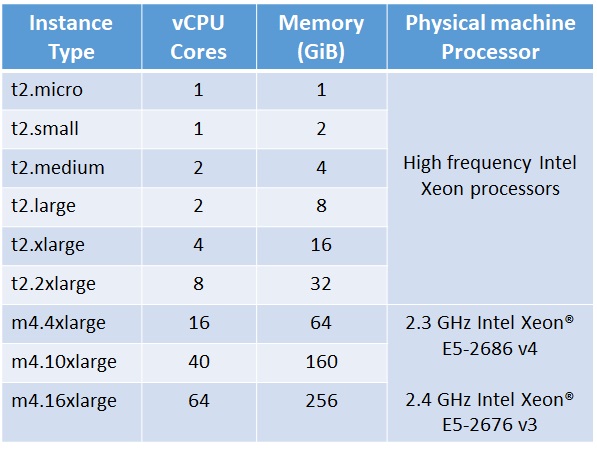}
	\end{center}
	\caption{EC2 General-Purpose Instances \cite{AmazonEC2Instances2021}}
	\label{Fig:AmazonEC2Instances}
\end{figure}

In this Section, we study of the costs (prices) of the instances (i.e., VM) offered by Amazon EC2 with respect to the amount of resources, location, and performance. The outcome of this study is leveraged later  while developing our SFC translation and mapping solutions to carefully place the instances in the physical infrastructure.
In particular, we considered Amazon EC2 General-purpose virtual machine instances of type T2 \cite{AmazonEC2Instances2021} which provide a large range of flavors as shown in~Fig~\ref{Fig:AmazonEC2Instances}. These general-purpose instances offer compute, memory and networking resources that can be used for diverse types of workloads including network functions. Each instance flavor defines the amount of vCPU and memory of the virtual machine and has a different cost. 
The table shows the hardware characteristics on which the VM flavor would run according to~Amazon (Fig.~\ref{Fig:AmazonEC2Instances}).

In the following, we study several aspects related to the cost of these flavors and their performance. The study includes an analysis of the instance cost versus its location, its software stack, and its allocated resource amount, and, finally, the VNF performance (i.e., packet processing capacity) versus the instance type.\\

\begin{figure}
	\begin{center} 
	\includegraphics[scale=.45]{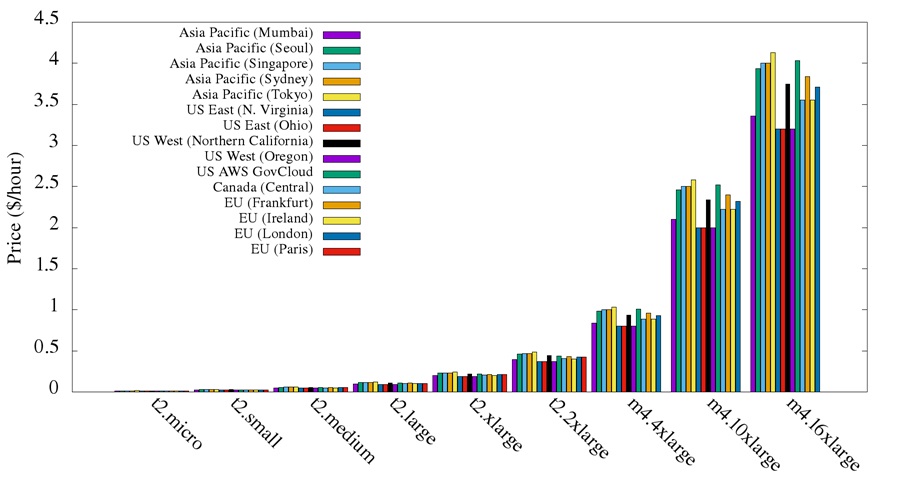}
	\end{center}
	\caption{Instance price for different locations}
	\label{Fig:InstancePriceVsLocations}
\end{figure}

$\bullet$ \textbf{Instance Cost vs. Location:}
Fig.~\ref{Fig:InstancePriceVsLocations} shows the instance price for~15~locations in the Amazon infrastructure. It is clear that instance costs significantly vary from one location to another. 
According to the figure, the difference in cost between two locations for the same instance type can go from 0.01\$ and can reach 1\$ for large instances.
It is worth noting that even a small difference in the instance cost may lead to a high impact even small difference in the instance cost. For instance,  0.1 \$/hour cost difference would translate into 86 million dollars a year considering 100K instances, and into around 2 billion dollars for around 2 Million instances, which is a lower-bound estimation of~the~number of~instances running on the Amazon EC2 infrastructre~\cite{NumberInstAmazon}. 

\begin{figure}
	\begin{center} 
	\includegraphics[scale=.5]{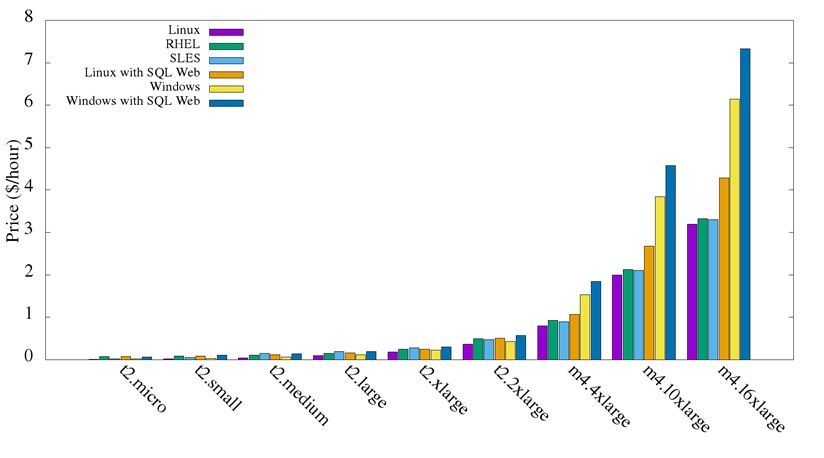}
	\end{center}
	\caption{Instance price for different software stacks (Oregon)}
	\label{CostVsSoftwareStack}
\end{figure}

$\bullet$ \textbf{Cost vs. Software Stack:}
Fig.~\ref{CostVsSoftwareStack} shows the price of different instances with different software stack. It considers only the instances located at Amazon AWS Oregon region. The figure shows that instance costs vary depending on the software stack. Linux distributions (e.g., Linux, RHEL, SLES) have similar costs and are much less expensive than instances running Microsoft Windows. Furthermore, adding additional software to the instance (e.g., SQL Web) would significanlty increase the price (e.g., up to 1.5\$ for large instances). \\

\begin{figure}
	\begin{center} 
	\includegraphics[scale=.5]{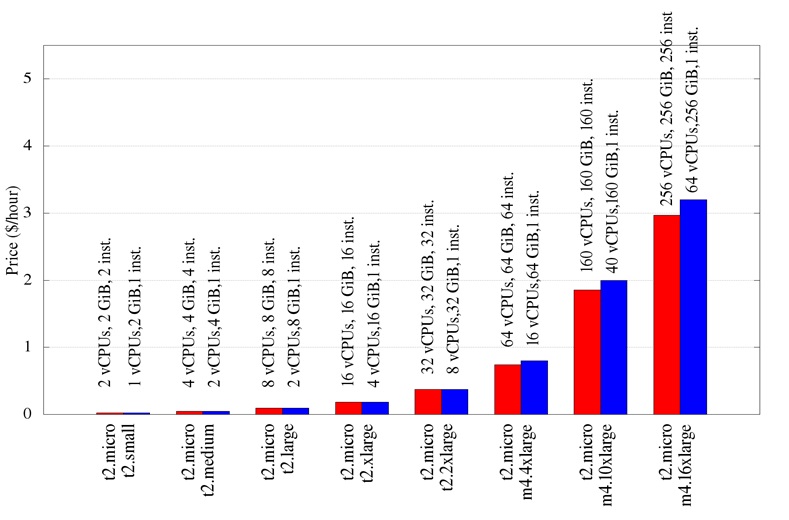}
	\end{center}
	\caption{price and size of Amazon EC2 isntances compared to t2.tiny instances (US West, Oregon)}
	\label{fig:InstanceVsPriceUSWest}
\end{figure}

$\bullet$ \textbf{Instance size vs. Cost}: The instance size refers to the amount of resource in terms of CPU and memory that an instance has. We hence aim at evaluating, for the same cost, how much resources we can provision when we use micro instances (.e., the smallest instance that has only 1 vCPU and 1GiB of memory) versus larger instances. To do so,  Fig.~\ref{fig:InstanceVsPriceUSWest} shows the price of all AWS instances and the the amount of resources they provide. It also shows how many t2.micro instances could be provisioned for the same price. 

For instance, as shown in the figure, the price of one m3.16xlarge instance (containing 64 vCPU and 256 GiB of memory) is 3.2 \$/hour. For almost the same price, one could provision 256 t2.micro instances offering 256 vCPU and 256 GiB of memory. This means that if provision 256 t2.micro services, we can get 192 (i.e.,~256-64) more vCPUs with the same amount of memory (256 GiB) compared to a single m3.16xlarge instance. The same note apply for the other types of instances.

As a result,  we can conclude that small instances are more cost-effective compared to large instances as, for the same cost, micro instances would provide roughly four times more vCPUs. This is of course interesting if the function/application could run normally in a distributed manner on several instances.
%

$\bullet$ \textbf{VNF Performance vs. Instance Type}
In this experiment, we try to evaluate the packet processing capacity of each instance, i.e.,~how much packets an instance could process when running a specific VNF. To do so, we conduct experiments using  different VNF types running on different instances while gradually increasing the packet arrival rate in order to assess the limit of the instance processing capacity. We assume that the processing capacity of the instance is reached when the CPU utilization of the instance reaches  90\% and the packet loss reaches 10\%.

\begin{figure}
	\begin{center} 
	\includegraphics[scale=.65]{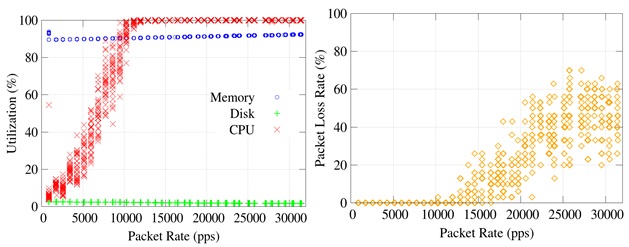}
	\end{center}
	\caption{Firewall [software: Shorewall, instance: t2.micro]}
	\label{Fig:FirewallShorewall_t2_micro}
\end{figure}

For instance, Fig~\ref{Fig:FirewallShorewall_t2_micro} shows how the utilization and the packet loss ratio evolve while increasing the incoming packet rate  for an Amazon Ec2 instance of type t2.micro running a Firewall (Shorewall Firewall \cite{Shorewall17}). We can see that the CPU utilization reaches 90\% and we start having packet loss when the incoming packet rate is around 10,000 packets per second (pps). This means that the processing capacity for this particular network function (Shorewall Firewall) on a t2.micro instance is around 10,000~pps. 

\begin{figure}
	\begin{center} 
	\includegraphics[scale=.65]{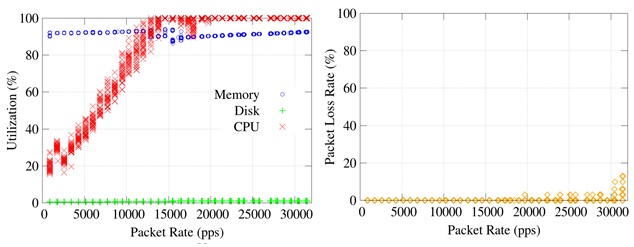}
	\end{center}
	\caption{IDS [software: Snort, instance: t2.micro]}
	\label{Fig:IDS_Snort_t2_micro}
\end{figure}

We have also conducted the same experiment while running the Snort Intrusion Detection System (IDS) \cite{Snort17} on the same type of instance. The results are reported in Fig.~\ref{Fig:IDS_Snort_t2_micro} and show that the processing capacity of the t2.micro instance running the Snort IDS function is 13000~pps.

\begin{figure}
	\begin{center} 
	\includegraphics[scale=.65]{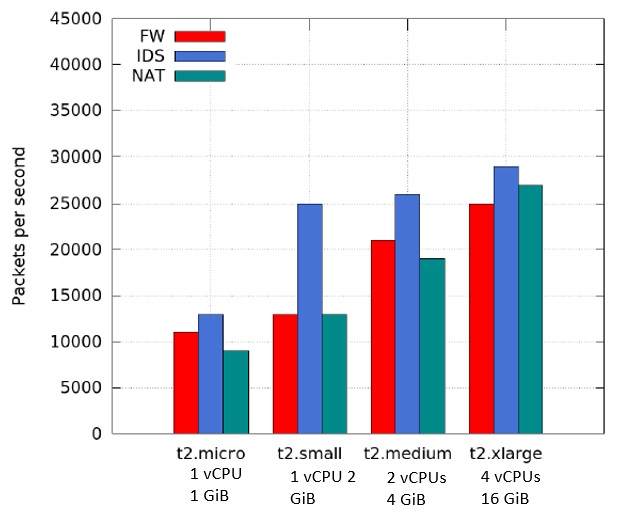}
	\end{center}
	\caption{VNF processing capacity per instance type}
	\label{Fig:VNFprocessingCapacityPerInstanceType}
\end{figure}

Fig.~\ref{Fig:VNFprocessingCapacityPerInstanceType} summarize the results for three types of network functions, namely a Firewall, and IDS and a NAT, that are running on different types of instances. We can clearly see in the figure that, for the same instance type, the packet processing capacity varies from one type of network function to another. Moreover, we can also notice that the processing capacity is not always proportional to the amount of allocated resources. Indeed, we can see in the figure that the t2.xlarge instance has 4 times resources that the t2.micro instance but is not able to process 4 times the amount of packets processed by the t2.micro. As a result, 4 micro instances would process much more packets than a single t2.xlarge instance. While it is not possible to provide a straightfoward explanation of this result (as we do not have access to internal statistics of the AWS infrastructure), the reasons of such result might be the network bottlenecks and also the heterogeneity of physical machines on which the instances are running (see~Fig.~\ref{Fig:AmazonEC2Instances}).

The above observation means that distributing a function over multiple small instances would allow higher packet processing capacity and also a lower cost according to the comparison reported in Fig.~\ref{fig:InstanceVsPriceUSWest}.\\


$\bullet$ \textbf{Study Outcomes:} 
we can summarize the study outcomes as follows:
\begin{itemize}
    \item Instance costs vary significantly from one location to another
    \item The software stack has a big impact on the instance cost
    \item The VNF processing capacity is not necessary proportional to the amount of resources
    \item The VNF processing capacity varies significantly from one function to another
     \item Small instances are more cost-effective and hence, if there is no synchronization cost, multiple instance deployment is more cost-effective and provides higher processing capacity.
\end{itemize}

Taking into consideration the above outcome, it is of utmost importance to  develop SFC provisioning solutions that are able to find the best tradeoff between cost (including instance price, synchronization and bandwidth costs) and~processing capacity.
In the following, we propose an Integer Linear Program to solve the SFC mapping phase and Two greedy solutions to deal with large-scale instances of the problem.
\section{Mapping Phase: Problem Formulation}\label{Section:ProblemFormulation}
In this section, we formulate the SFC mapping problem  as an Integer Linear Program (ILP) with the objective of minimizing the SFC provider's operational costs in terms of instance deployment costs,  bandwidth and  synchronization costs. 

\begin{table}[]
\centering
\caption{Table of notations}
\label{TableofDefinitions}
\begin{tabular}{|c|l|}
\hline
\textbf{Symbol} & \multicolumn{1}{c|}{\textbf{Definition}} \\ \hline
${G}=({N},{P})$ & Graph $G$ where $N$ is the set of nodes \\ &and $P$ is set of physical links  \\ \hline
$V=({I},{L})$ & The virtual network Graph $V$ with $I$ is the set of VNF \\ instances & and $L$ is the set of virtual links  \\ \hline
$C_{n}$ & Available capacity at POP $n \in N$ expressed in number\\ &of instances\\ \hline
$B_{{m}{n}}$ &  Bandwidth capacity of the physical link connecting \\ & nodes $m$ and $n$ \\ \hline
$b_{{i},{j}}$ & Bandwidth requirement of the virtual link \\ & connecting instances $i$ and $j$ \\ \hline
$\delta_{{i}{m}}$ & Deployment costs per unit of time  for 
\\ & VNF instance $i$ into POP $m$ \\ \hline
$\Delta_{m,n}$ & Bandwidth cost per bandwidth unit in physical  \\ & link $({m},{n})$ \\ \hline
$f_{{i}{m}}$ & Boolean constant set to 1 if VNF instance $i$ has \\ & to be embedded into node $m$ \\ \hline
$s_{ij}$& Boolean constant set to 1 if there is a synchonization \\ & between instances $i$ and $j$ \\
\hline
$x_{{i}{m}}$ & Boolean decision variable indicating whether or not \\ & instance $i$ is embedded into node $m$ \\ \hline
$y_{{i}{j},{m}{n}}$ &  Boolean decision variable indicating whether \\ & virtual link $({i}, {j})$ is mapped into physical link $({m}, {n})$ \\ \hline
$\mathbb{C}$ & Operational cost  \\ \hline
$\mathbb{S}$ & Synchronization cost \\ \hline

\end{tabular}
\end{table}


\begin{figure}[ht]
	\centering
		\fbox{\includegraphics[width=0.45\textwidth]{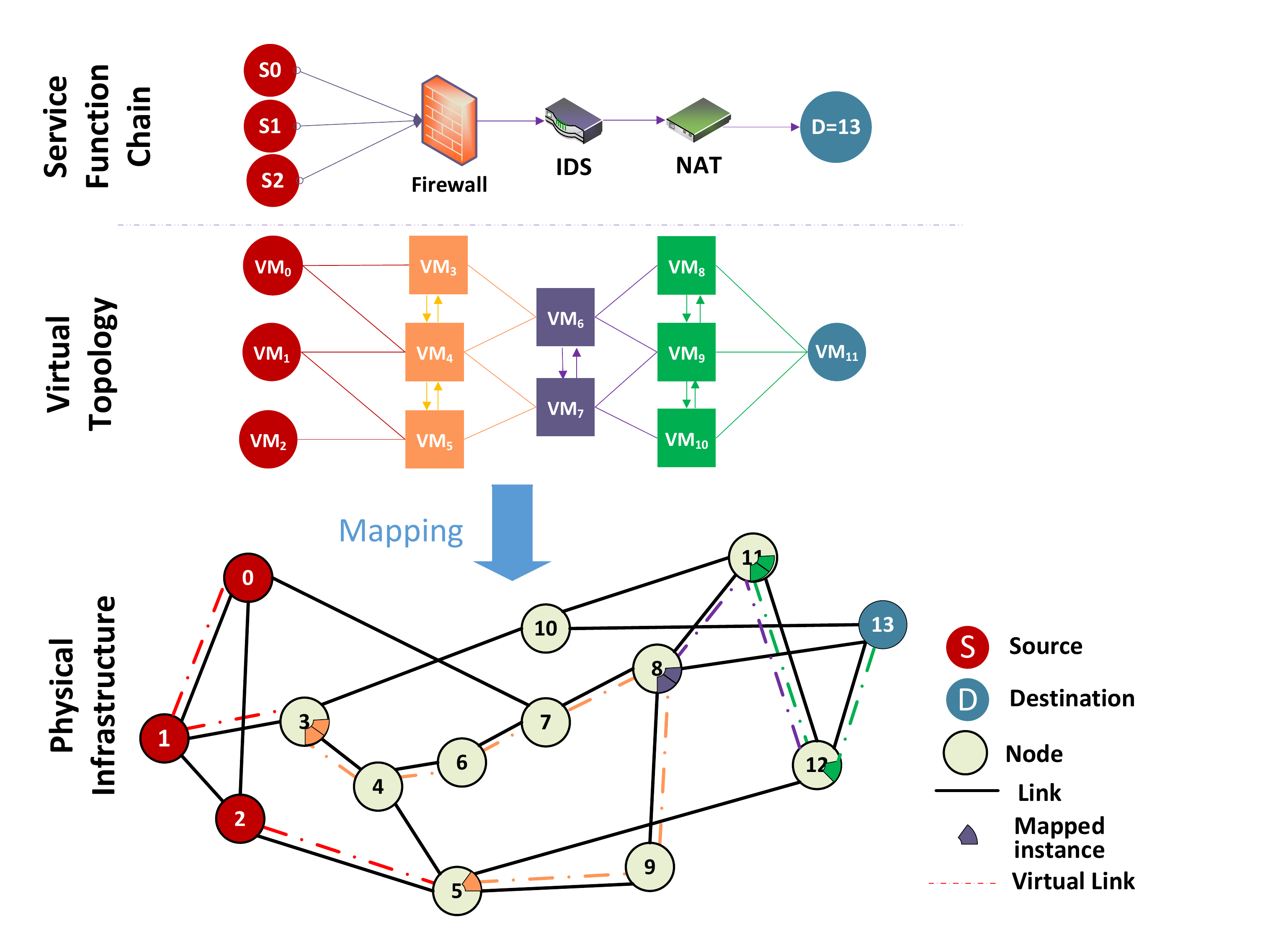}}
	\caption{SFC embedding problem}
	\label{fig:VNF-SFC_Embedding}
\end{figure}

The physical infrastructure owned by the SFC provider is made out from several POPs that are geographically distributed. The infrastructure is modeled by a  graph ${G}=({N},{P})$ where $N=\{0,1,..., |N|\}$ represents the set of POPs and $P=\{(m,n) \in (N \times N)\ |\ \text{$m$ and $n$ are directly connected}  \}$ denotes the set of physical links that connect the POPs. Each POP ${n}\in{N}$ contains an amount of physical resources  $C_{n}$ expressed  as the maximal number of t2.tiny instances that the POP can host. Note that a t2.tiny instance contains 1 vCPU, 1 GiB and 1 GB of memory and disk, respectively.
A physical link $(m,n) \in P$ that connects the POP $m$ with POP $n$ has a bandwidth capacity $B_{{m}{n}}$.

Furthermore, a service function chain is represented as a graph $V=(I, L)$ where $I=\{0,1,..., |I|\}$ is the set of virtual instances in the chain and $L$ is the set of virtual links connecting them.

Each VNF instance ${i} \in I$ has a resource requirement of 1 vCPU, 1 GiB of memory, and 1GB of storage. Each virtual link $(i,j)\in L$ has bandwidth requirement $b_{ij}$. 
It is worth noting that, for simplicity, the endpoints of the chain (i.e., sources and destinations) are considered also instances with requested resources equal to zero. They are constrained to be mapped onto particular physical POPs that are provided in the VNF request.

Furthermore, we define two decision variables. The first one is denoted as $x_{{i}{m}} \in \left \{ 0,1 \right \}$ and indicates whether or not VNF instance ${i}$ is embedded  into POP $m$. 
The second decision variable is denoted as $y_{{i}{j},{m}{n}} \in \left \{ 0,1 \right \}$. If $y_{{i}{j},{m}{n}}=1$, the virtual link $(i,j)$ uses the physical link ${m}{n}$. It is worth noting that a virtual link is embedded through a physical path (i.e, multiple connected physical links). Hence, several physical links could be used to embed a virtual link. In other words, if $y_{{i}{j},{m}{n}}=1$, the physical link $(m,n)$ is part of the physical path used to embed the virtual link $(i,j)$.

\textbf{$\bullet$ Objective Function:} The objective function when embedding an SFC request aims at minimizing the operational costs $\mathbb{C}$ and synchronization cost of the embedded VNF instances $\mathbb{S}$. It can be expressed as:
\begin{equation}\label{eq:ObjFctn}
J = \min_{\substack{(x_{{i}{m}})_{i\in I, m\in N}\\ (y_{{i}{j},{m}{n}})_{(i,j)\in L, (m,n)\in P}}} \left ( \mathbb{C} + \mathbb{S} \right) 
\end{equation}

In the following, we provide more details on how to compute the operational and synchronization costs:

\textbf{$\bullet$ Synchronization cost:} the synchronization cost can be expressed as following: 

\begin{equation}\label{eq:SynchCost}
\mathbb{S}=  \sum_{{(i,j)}\in{L}} \sum_{{(m,n)}\in{P}} y_{{i}{j},{m}{n}} s_{ij} \hspace*{1mm}b_{{i}{j}}\hspace*{1mm} \Delta_{{m}{n}} 
\end{equation}
where $y_{{i}{j},{m}{n}}$ indicates whether or not the physical link $(m,n)$ is used for embedding the virtual link $(i,j)$.  The Boolean variable $s_{ij}$ is equal to 1 if instances $i$ and $j$ implement the same VNF type and hence require synchronization among them to operate. In this case, there is a synchronization cost computed as $b_{{i}{j}}$ $\Delta_{{m}{n}}$, which is the cost of using bandwidth needed to exchange synchronization data between $i$ and $j$.  

\textbf{$\bullet$ Instance and link operational cost:} it is the cost of running a the VNF instances on the infrastructure and the bandwidth consumed by the virtual links connecting them. It can be expressed as follows: 


\begin{equation} \label{eq:opcosts}
\begin{split}
\mathbb{C} &= \sum_{{m}\in{M}} \sum_{{i}\in{I}} x_{{{i}}{m}}\hspace*{1mm} \delta_{{i}{m}}\hspace*{1mm} \\
& + \hspace*{1mm}\sum_{{(i,j)}\in{L}} \sum_{{(m,n)}\in{P}} y_{{i}{j},{m}{n}} \hspace*{1mm} (1-s_{ij}) \hspace*{1mm}b_{{i}{j}} \hspace*{1mm} \Delta_{m,n} 
\end{split}
\end{equation}
where $\delta_{{i}{m}}$ is the deployment cost (expressed in dollars per unit of time) of VNF instance $i$ into POP $m$. It is worth noting that $\delta_{{i}{m}}$ varies from one POP to another as it depends on several factors including the electricity price in the POP, the type of the VNF, the license, and the operating system as suggested by the conducted Amazon EC2 study. The first term of the equation (Eq.~\ref{eq:opcosts}) represents the total cost of deploying the VNF instances. The second term of the operational costs is the total cost of bandwidth consumed by the virtual links.  $\Delta_{m,n}$ denotes the cost in dollars (per bandwidth unit and unit of time) for the physical link $(m,n)$. 

The above objective function is subject to the following set~of~constraints:

\textbf{$\bullet$ SFC Endpoints embedding constraint:} 
SFC endpoints (the sources and the destinations) should be embedded into specific POPs stated in the request.We define the boolean variable $f_{{i}{m}}$ (provided as an input to the ILP) that is equal 1 when the instance $i$ is an endpoint that has to be embedded in POP $m$. The following equation captures this constraint:
\begin{equation} \label{eq:csrtEndPoints}
x_{{i}{m}} \geq f_{{i}{m}} \ \  \ \ \forall {m \in N},\ \forall {i \in I}
\end{equation}

\textbf{Instance embedding constraint:}  This constraints ensures that each VNF instance $i$ is embedded one and only once. It can be expressed as:
\begin{equation}\label{eq:csrtInstembed}
\sum_{{m} \in {N}} x_{{i}{m}} = 1 \hspace*{7mm}  \forall {i \in I}
\end{equation}

\textbf{$\bullet$ Resource capacity constraint: } 
this constraint ensures that any hosting POP has enough resources to host the VNF instances.
%
\begin{equation}\label{eq:csrtCap}
\sum_{{i} \in I} x_{{i}{m}} \leq C_{m} \hspace*{7mm} \hspace*{1mm} \forall {{m} \in {N}}
\end{equation}%
where $C_{m}$ represents the available capacity at POP $m$.

\textbf{$\bullet$ Bandwidth constraint:} 
we must also ensure that the bandwidth capacity required to embed all virtual links in a physical link does not exceed its available bandwidth. This can be expressed as follows: 
\begin{equation}\label{eq:csrtBW}
\sum_{i,j \in L} y_{{{i}{j}},{{m}{n}}} \hspace*{1mm}b_{{i}{j}}\leq B_{{m}{n}} \hspace*{5mm} \forall{\ (m,n) \in P} 
\end{equation}%

\textbf{$\bullet$ Flow conservation constraint:} we must also ensure that the incoming traffic to a physical node is equal to its outgoing traffic unless this POP is a~source or a~destination. 
This constraint can be expressed as: 
\begin{equation} \label{eq:csrtFlowCons2}
\begin{split}
\sum_{{(n,m)}\in P} \sum_{{(i,j)}\in L} y_{{i}{j},{n}{m}} \hspace*{1mm} b_{{i}{j}}
 - \sum_{{(i,j)}\in L} x_{{j}{m}}\hspace*{1mm} b_{{i}{j}}  &\\
= \sum_{{(m,n)}\in P} \sum_{{(i,j)}\in L} y_{{i}{j},{m}{n}} \hspace*{1mm} b_{{i}{j}}
 - \sum_{{(i,j)}\in L} x_{{i}{m}}\hspace*{1mm} b_{{i}{j}} &
 \ \ \ \  \forall{\ m \in N}
\end{split}
\end{equation}
%
%
The service chain embedding problem is an \emph{NP}-hard problem as it generalizes bin-packing problem; therefore, finding an optimal solution is not viable due to the large number of requests processed in the production environment. Hence, we propose two heuristics in the following Section to solve this problem and~explore potential solutions.

\section{Mapping Phase - Proposed Solutions}\label{Section:ProposedSolutions}
In this Section, we address the \emph{NP}-hardness of the problem by putting forward two heuristics solutions, a Baseline algorithm and a more sophisticated algorithm called SFC decompoSition-based ProvisIoNing (SPIN). Both solutions assume multiple sources and a single destination to simplify the problem and aim at mimimizing SFC provider's operational and synchronization costs while ensuring that accepted requests satisfy their end-to-end delay requirement.  In the following, we provide more details about the two algorithms.

\subsection{Solution 1: Baseline Algorithm}
The baseline algorithm is a intuitive algorithm that aims satisfying the requirements of the SFC in terms of resources (e.g., CPU, memory, bandwidth) and end-to-end delay while minimizing instances costs.  The algorithm proceeds with the following steps. 

The first Step is to estimate the number of instances and virtual links required for the whole chain. The number of instances is simply equal to the number of t2.micro instances needed to process the arriving packet rate. The processing capacity of a t2.micro instance is estimated using the technique described in Section~\ref{Section:StudyAWS} (e.g.,~Fig.~\ref{Fig:FirewallShorewall_t2_micro} and~Fig.~\ref{Fig:IDS_Snort_t2_micro}). Once the number of instances for each VNF is estimated, the virtual topology is built.

The second step is to allocated resources for this virtual topology as shown in~Algorithm~\ref{alg:Baseline}. For each source instance of the virtual topology, we start by embedding the virtual nodes (i.e.,~instances) connected it to it (i.e.,~neighbors). For each of these instances, we recursively embed its neighbors by calling recursively the function $EmbedNeighbors(instance\ i)$ (Algorithm~\ref{alg:EmbedNeighbors}).

The complexity of this recursive algorithm is $O(|I|^2)$ where $|I|$ is the number of virtual instances in the virtual topology.
\begin{algorithm}
\caption{Baseline} \label{alg:Baseline}
\begin{algorithmic}
\STATE \textbf{Input:} Virtual topology V= (I,L)
\STATE \textbf{Input:} Placement constraint $(f_{im})_{i\in I, m\in N}$\\
\STATE \textbf{Input:} Virtual Topology Destination $d \in N$
\STATE \textbf{Output:} Boolean Embedded  
\FORALL{$i \in I$ such that $i$ is a source (i.e., $\sum_{m\in N} f_{im}=1$)}
\STATE $s \Leftarrow$ the hosting physical node of source instance $i$ (i.e.,$f_{is}=1$)  
\STATE $x_{is} \Leftarrow 1$  
\STATE Return EmbedNeighbors(i)
\ENDFOR
\end{algorithmic}
\end{algorithm}
\begin{algorithm}
\caption{EmbedNeighbors(instance i)} \label{alg:EmbedNeighbors}
\begin{algorithmic}
\STATE $s \Leftarrow$ Physical node hosting instance $i$ 
\FORALL{$j \in neighbors(i)$ (Embedding instances connected to $i$) }
    \IF{$j$ is not embedded}
            \STATE Find $m$ such that $m \in ShortestPath(s, d)$ \& $C_m \geq 1$ \& $ PathBandwidth(s,m)\geq b_{i,j}$
            \IF{$m$ exists}
                \STATE $x_{jm} \Leftarrow 1$   (Embed $j$ in $m$)
                \STATE $y_{{i}{j},{s}{m}} \Leftarrow 1$   (Embed virtual link $(i,j)$ in physical path $(s,m)$) 
                \STATE $C_m \Leftarrow C_m -1$ (Update the node capacity) 
            \ELSE
                \STATE Return False (Instance $j$ is not embeddable)
            \ENDIF
    \ENDIF
\ENDFOR
\FORALL{$j \in neighbors(s)$}
    \IF{$j$ is not embedded}
        \STATE Return EmbedNeighbors($j$) (Embedding instances connected to $j$)
    \ENDIF
\ENDFOR
\STATE Return True (all instances were embedded)
\end{algorithmic}
\end{algorithm}
 
\subsection{Solution 2: SFC decompoSition-based ProvisIoNing (SPIN) Algorithm}
This algorithm is called SFC decompoSition-based ProvisIoNing (SPIN) and proceeds into four phases (Algorithm \ref{alg:SPIN}). In the first phase, we  estimate the number of instances for each VNF and estimate the number of virtual links just like the way it is done by the Baseline Algorithm. The second phase is the decomposition phase where the virtual topology is divided into subchains where each subchain is a chain of VNF instances that contains a single instance of each VNF type and connects one source to one destination. 

The third phase is the subchain embedding Phase (Algorithm~\ref{alg:EmbedSubchain}) where each subchain is embedded in the shortest path between the source and destination of the subchain denoted as $P$. The path $P$ is the selected as the one with lowest cost and that has a delay satisfying the e2e delay requirement of the chain and has enough resources to embed the subchain ($FreeInst(P)$ is the number of free instances in the  path $P$ and  $NumberInstances(SC_k)$ is the number of instances needed by subchain $SC_k$). The virtual links intended to carry the synchronization traffic are then provisioned between the same-type VNF instances (Function $EmbedSynchronizationVirtualLinks(V)$). 

The last phase is the optimization phase  \ref{alg:Optimization} that consists of selecting each instance and explore the possibility of migrating it in one of physical nodes that are neighboring its current physical location. The goal is to  further reduce operational and synchronization costs (Eq.~\ref{eq:ObjFctn}) while always ensuring that the requested bandwidth and e2e delay is satisfied. 

The complexity of SPIN algorithm is $O(K)$ where $K$ is the number of subchains. The complexity of the optimization phase is $O(|V|)$ where $V$ is the number of virtual instances in the virtual topology.

\begin{algorithm}
\caption{SPIN} \label{alg:SPIN}
\begin{algorithmic}
\STATE \textbf{Input:} Virtual topology $V= (I,L)$
\STATE \textbf{Input:} Placement constraint $(f_{im})_{i\in I, m\in N}$\\
\STATE \textbf{Input:} Virtual Topology Destination $d \in N$
\STATE \textbf{Output:} Boolean $Embedded$, $VLEmbedded$
\STATE Decompose $V$ into $K$ subchains $(SC_k)_{(k=1..K)}$ 
\REPEAT
\STATE $Embedded \Leftarrow EmbedSubchain(SC_k)$
\STATE $k \Leftarrow k +1$ 
\UNTIL{$Embedded=False$ $\|$ $\ k=K+1$}
\STATE $VLEmbedded \Leftarrow EmbedSynchronizationVirtualLinks(V)$
\IF{($Embedded$ \& $VLEmbedded$)=True (Embedding is succesful)}
    \STATE Optimization($V$) (optimization phase)
    \STATE Return $True$
\ELSE
    \STATE Return $False$
\ENDIF
\end{algorithmic}
\end{algorithm}
\begin{algorithm}
\caption{EmbedSubchain(subchain $SC_k$)} \label{alg:EmbedSubchain}
\begin{algorithmic}
\STATE  $P \Leftarrow$ Find path with minimal cost such that $delay(P)\leq delay(SC_k$ \& $FreeInst(P) \leq NumberInstances(SC_k)$ \& $Bandwidth(P)\geq Bandwidth (SC_k)$
\IF{$P$ exists}
    \STATE Embed $SC_k$ in $P$
    \STATE Return True ($SC_k$ is successfully embedded)
    \ELSE
    \STATE Return False ($SC_k$ is not embeddable)
\ENDIF
\end{algorithmic}
\end{algorithm}
\begin{algorithm}
\caption{Optimization(VirtualTopology $V$)} \label{alg:Optimization}
\begin{algorithmic}
\FORALL{$i \in V$ (Parse all instances)} 
    \STATE $n \Leftarrow$ Physical node hosting $i$
    \FORALL{$m \in neighbors(n)$ (Explore migrating $i$ to neighboring nodes) }
        \STATE $Cost\Leftarrow CurrentVCost(V)$ (Compute Embedding Cost Eq.~\ref{eq:ObjFctn})
        \STATE $NewCost\Leftarrow VCost(V,i,m)$ (Compute cost assuming $i$ is hosted in $m$)
        \IF{$CheckConstraints(V)$ \& $NewCost < Cost$  (all resource constraints should be satisfied)}
            \STATE Migrate($i$, $m$) (migrate instance $i$ to physical node $m$
        \ENDIF
    \ENDFOR
\ENDFOR
\end{algorithmic}
\end{algorithm}
\section{Performance Evaluation}\label{Section:PerformanceEvaluation}
\subsection{Simulation Setup}
In order to evaluate the performance of the proposed algorithms, we developed C-based simulator that simulates the physical topology and carry out the translation and the mapping of the SFC requests. Each simulation assumes the arrival of requests during two months. 

The physical infrastructure is assumed to contain 25 nodes with each node having a hosting capacity randomly set between 50 and 100 t2.micro instances. The nodes are connected with 10 Gbps links  with propagation delays randomly set between 10 and 50~ms.

The SFC were generated randomly with an an average arrival rate set between 0.1 and 0.15 rps following a Poisson distribution. The average lifetime of the requests follows an exponential distribution with an average of 1 hour. The average number of VNFs per SFC is 10 with an average number of source around 7. 
The demand in terms of packet arrival for each SFC is generated randomly between 2000 and 120 000 packet per seconds. 

The end-to-end delay requirement of an SFC request is computed as follows   $\max_{s,d} (t_{s,d})~\times ~130\%$ where $t_{s,d}$ is the path latency between a source $s$ and a destination $d$ where $s$ and $d$ are a source and a destination of the SFC request. This ensures that, theoretically, the end-to-end delay requirement between the sources of the SFC and its destination could be satisfied as they it is 30\% higher than any path between the sources and the destinations of the request. 

We also assume that we have 9 types of VNFs. The packet processing capacity of each type of VNF is generated randomly between 2000 to 12000 packet per second (pps) when  running on t2.micro instance.
The synchronization cost among same-type instances is 0.01\$/hour multiplied by the type of the instance.
We used the Amazon EC2 instance prices as instance costs. The instance revenue as the  cost of the instance plus 0.1\$/hour. this means that there is 0.1\$/hour profit for the SFC provider each instance.

In the next subsection, we present the results generated for the two proposed algorithms under the above-described simulation setup.
\subsection{Simulation Results}
We first compare the performance of the Baseline and SPIN algorithms for an arrival rate 0.03 request per second (rps).

\begin{figure}
	\begin{center} 
	\includegraphics[scale=.80]{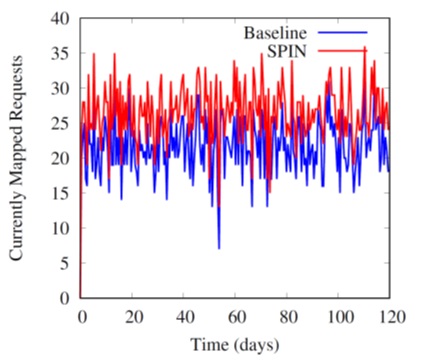}
	\end{center}
	\caption{Number of mapped requests over time (request arrival rate 0.03 rps)}
	\label{Fig:NumberMappedRequestsOverTime}
\end{figure}

\begin{figure}
	\begin{center} 
	\includegraphics[scale=.80]{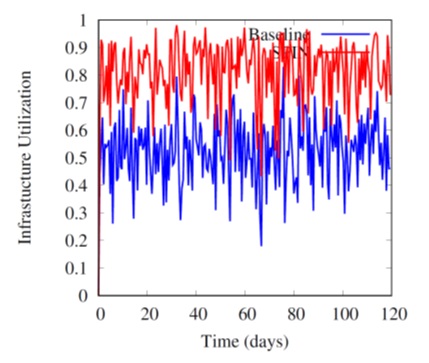}
	\end{center}
	\caption{Infratructure CPU utilization over time (request arrival rate 0.03 rps)}
	\label{Fig:InfratructurUtilizationOverTime}
\end{figure}

As shown in Figures~\ref{Fig:NumberMappedRequestsOverTime} and~\ref{Fig:InfratructurUtilizationOverTime},  SPIN maps 25\% more requests and leads to around 37\% higher CPU utilization in the whole infrasturcture.

To further assess the performance of the two proposed greedy algorithms for different scenarios, we computed the following metrics while varying the SFC requests arriving rate: 
\begin{itemize}
    \item Acceptance Ratio: it is computed as the ratio of the number of accepted SFCs to the total number of received SFC requests. Accepted requests refers to the ones for which the algorithm succeeded to found enough resources for the SFC while satisfying its e2e delay requirements.
    \item Infrastructure utilization: amount of used CPU resource divided by the total available resource (CPU)
    \item Cumulative Profit: is is computed as the revenue of the SFC provider minus its operational costs including the costs of the instance, bandwidth, and synchronization. The cumulative profit is computed for the duration of the experiment.
    \item Average end-to-end (e2e) delay: the average end-to-end delay between the sources and destinations of the SFC requests that were successfully embedded throughout the experiment.
\end{itemize}

In the following, we provide and discuss the obtained results for each metric.

\begin{figure}
	\begin{center} 
	\includegraphics[scale=.70]{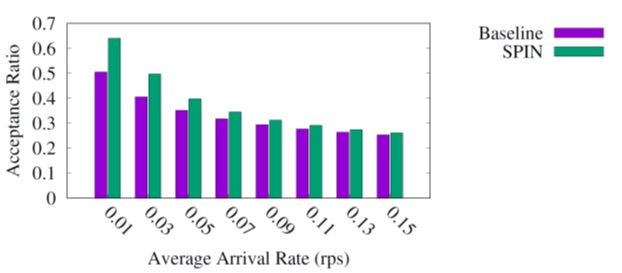}
	\end{center}
	\caption{Acceptance Ratio}
	\label{Fig:AcceptanceRatio}
\end{figure}
\begin{figure} [htp]
	\begin{center} 
	\includegraphics[scale=.70]{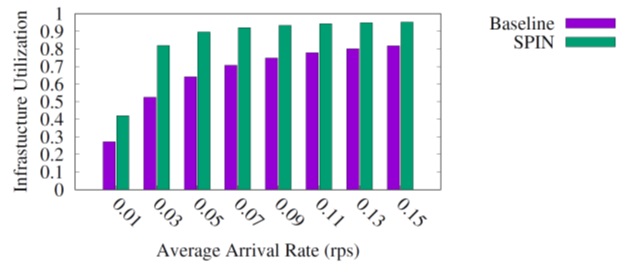}
	\end{center}
	\caption{Infrastructure Utilization}
	\label{Fig:InfrastructureUtilization}
\end{figure}
The first considered metric is the acceptance ratio and is depicted in~Fig.~\ref{Fig:AcceptanceRatio}. The figure shows that, even a low request arrival rate, the baseline fails to accommodate 50\% of the requests where as SPIN succeeds to accommodate up to 65\% of the requests. This means that even if the infrastructure's utilization is low and resources are available (Fig.~\ref{Fig:InfrastructureUtilization}), the baseline, unlike SPIN, is not able to efficiently leverage such available resources.  
As the arrival rate is increased, the acceptance ratio goes down for both algorithms as the infrastructure becomes saturated as Fig.~\ref{Fig:InfrastructureUtilization}. However, SPIN still outperforms the baseline in terms of acceptance ratio.


Furthermore, as illustrated in~Fig.~\ref{Fig:InfrastructureUtilization}, SPIN accepts up to 25\% more requests for low arrival rates, showing that it  allows to efficiently leverage the infrastructure resources compared to the baseline. It is worth noting that for high arrival rates SPIN succeeds to reach 95\% utilization compared to 80\% utilization for the Baseline.


\begin{figure}
	\begin{center} 
	\includegraphics[scale=.70]{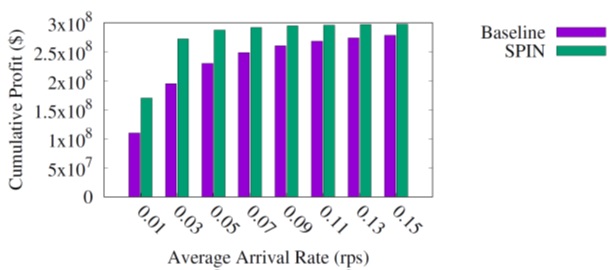}
	\end{center}
	\caption{Cumulative Profit}
	\label{Fig:CumulativeProfit}
\end{figure}

We also study cumulative profit generated by each of the two algorithms (Fig.~\ref{Fig:CumulativeProfit}). The figure shows that the profit generated by SPIN exceeds by up to 30\% the one generated by the baseline, especially for low arrival rates.

\begin{figure}
	\begin{center} 
	\includegraphics[scale=.70]{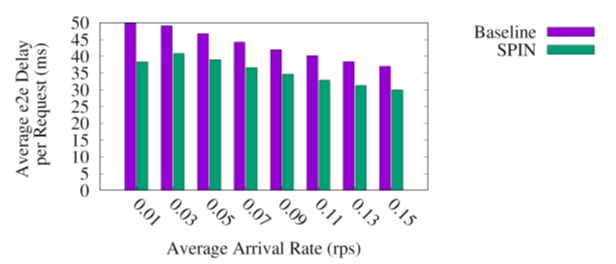}
	\end{center}
	\caption{Average e2e Delay per Request}
	\label{Fig:Averagee2eDelayPerRequest}
\end{figure}

Finally, Fig.~\ref{Fig:Averagee2eDelayPerRequest} shows the average end-to-end delay per request for the accepted SFCs. It~shows that during low utilization, SPIN  reduces by up to 35\% e2e delay and by up to 25\% the e2e delay for high arrival rates. This shows that SPIN does not only satisfy the  requests' requirements in terms of e2e delay  but further reduce it compared to the baseline.


\section{Conclusion and Future Directions}\label{Section:Conclusion}

Selecting the right placement for the VNF, the number and the type of the VM instance is a major challenge for cloud providers as it has a paramount impact not only on the performance but also the cost. In this paper, we started by studying these tradeoff using general-purpose Amazon EC2 instances. For instance, we found that Micro-instances (small instances) are more cost-effective. We also find that the performance of a virtual machine is not always proportional to the amount of resources that are allocated. Hence, we found that, provisioning several small instances, when the function could be distributed, would provide better performance than big instances with a smaller cost. 

Furthermore, we investigated profit-driven resource allocation by mathematically modeling the problem as an Integer Linear Program and proposing two heuristics, a baseline and a more sophisticated algorithm dubbed SPIN, which allows to improve the performance of the mapping with more accepted chains and hence to increase profits.

This work opens the door for more research opportunities. For instance, it~would be interesting to further develop VNF Benchmarks with the development of more sophisticated procedures to benchmark VNFs depending on the nature of the implemented network function. It is also of utmost importance to devise resource consumption models for specific VNFs taking into consideration the VNF characteristics and the hosting. Another research avenue is to assess synchronization costs among same-instance VNFs depending on the type of the function and the instance locations.

More work should also be done on the Management of VNFs by developing platform-aware resource allocation as the performance of a virtual machine significantly depends on the hosting platform (dedicated hardware versus software, server model, type and amount of resource).
\ifCLASSOPTIONcaptionsoff
  \newpage
\fi

\newpage
\begin{IEEEbiographynophoto}{Tarik Moufakir} 
received the M.Sc. in Computer Science in 2013 from UQAM University, Canada and another M.Sc. in Computer Science from ENST Bretagne, France in 2002. He is currently pursuing the Ph.D. degree in Information Technology at \'Ecole de Technologie Sup\'erieure (\'ETS) in Montreal, QC, Canada. His main research interests are software-defined networking and cloud computing.
\end{IEEEbiographynophoto}
\begin{IEEEbiographynophoto}{Mohamed~Faten~Zhani}
is Associate Professor with the Software and IT Engineering Department, \'Ecole de Technologie Sup\'erieure (\'ETS), Montreal, QC, Canada. His research interests include cloud computing, network function virtualization, software-defined networking and resource management.
\end{IEEEbiographynophoto}
\begin{IEEEbiographynophoto}{Abdelouahed~Gherbi}
received the Ph.D. degree in computer engineering from Concordia University, Canada. He is currently Associate Professor with the Software and IT Engineering Department, \'Ecole de Technologie Sup\'erieure (\'ETS), Montreal, QC, Canada. His research interests include model-driven software engineering, modeling and analysis techniques for real-time and critical software systems, software performance, high availability, and security.
\end{IEEEbiographynophoto}
\begin{IEEEbiographynophoto}{Moayad Aloqaily}
is an Assistant Professor with the Faculty of Engineering, AI Ain University, United Arab Emirates. He received his Ph.D. degree in electrical and computer engineering from the University of Ottawa, Ontario, Canada.His current research interests include the applications of AI and ML, connected and autonomous vehicles, blockchain solutions, and sustainable energy and data management.
\end{IEEEbiographynophoto}
\begin{IEEEbiographynophoto}{Nadir Ghrada}
received the M.Sc. in Computer Science in 2018 from \'Ecole de Technologie Sup\'erieure (\'ETS) in Montreal, QC, Canada. His main research interests are cloud computing.
\end{IEEEbiographynophoto}






\end{document}